\documentclass[aps,prev,twocolumn,preprintnumbers,floatfix,nofootinbib]{revtex4-1}
\pdfoutput=1
\usepackage{graphicx}
\usepackage{bm}
\usepackage{times}
\usepackage{slashed}
\usepackage{color}
\usepackage{aas_macros}
\usepackage{slashed}
\usepackage{lipsum}
\usepackage{subfigure}
\usepackage{multirow}
\usepackage{amsmath}
\usepackage{array} 
\usepackage{varwidth} 

\newcommand{\be}{\begin{equation}}
\newcommand{\ee}{\end{equation}}
\newcommand{\bea}{\begin{eqnarray}}
\newcommand{\eea}{\end{eqnarray}}

\newcommand{\sigmav}{\ensuremath{\langle \sigma v \rangle}}
\DeclareMathOperator*{\argmax}{arg\,max}

\begin{document}

\title{A Robust Method for Treating Astrophysical Mismodeling in \\ Dark Matter Annihilation Searches of Dwarf Spheroidal Galaxies}
\author{Tim Linden}
\email{linden.70@osu.edu}
\affiliation{Center for Cosmology and AstroParticle Physics (CCAPP), and \\ Department of Physics, The Ohio State University Columbus, OH, 43210 }

\begin{abstract}
\noindent Fermi-LAT observations have strongly constrained dark matter annihilation through the joint-likelihood analysis of dwarf spheroidal galaxies (dSphs). These constraints are expected to be robust because dSphs have measurable dark matter content and produce negligible astrophysical emission. However, each dSph is dim, with a predicted flux that typically falls below the accuracy of the background model. We show that this significantly diminishes the reliability of previous joint-likelihood algorithms, and develop an improved analysis that directly accounts for the effect of background mismodeling. This method produces more robust limits and detections of dark matter in both real and mock data. We calculate improved limits on the dark matter annihilation cross-section, which differ by nearly a factor of two from previous analyses --- despite examining identical data.
\end{abstract}

\maketitle

Over the last decade, Fermi-LAT observations have strongly constrained the dark matter annihilation cross-section. In many cases, these observations have met, or exceeded, the thermal annihilation cross-section --- at which dark matter particles in thermal equilibrium in the early universe would freeze out to produce the observed dark matter abundance~\citep{Steigman:2012nb}. Intriguingly, several excesses are consistent with a $\sim$30-100~GeV dark matter particle annihilating near this cross-section~\citep{Hooper:2012jc, Bertoni:2016hoh, Cuoco:2016eej, Cui:2016ppb, Cholis:2019ejx}, most notably the $\gamma$-ray excess observed surrounding the Milky Way's galactic center~\citep{Goodenough:2009gk, Abazajian:2012pn, Gordon:2013vta, Daylan:2014rsa, TheFermi-LAT:2015kwa, TheFermi-LAT:2017vmf}. These facts place us in a new ``precision" regime, where the accuracy of dark matter annihilation constraints is critical.

The most robust annihilation constraints stem from the Milky Way's dwarf spheroidal galaxies (dSphs)~\citep{GeringerSameth:2011iw, Ackermann:2011wa, GeringerSameth:2012sr, Ackermann:2013yva, Ackermann:2015zua, Fermi-LAT:2016uux, Hoof:2018hyn}. While dSphs are unlikely to be the brightest annihilation sources, three factors produce strong constraints. First, stellar dispersion measurements constrain their dark matter content~\citep{2011MNRAS.418.1526C, Bonnivard:2016tas, Strigari:2018utn}. Second, they lie far from the complex Milky Way plane. Third, they produce negligible astrophysical emission~\citep{Winter:2016wmy}. These factors are unique to dSphs and validate their status as the ``silver bullet" target for indirect detection.

However, there are two complications in setting annihilation constraints with dSphs. First, while their dark matter content is measured, their relative annihilation rate is uncertain due to variations in the dark matter radial profile, as well as effects such as asphericity and triaxiality. These complexities are translated into statistical errors in the dSph ``J-factor", calculated as the square of the dark matter density integrated over the line of sight. Significant efforts are underway to quantify and model these uncertainties~\citep{2011MNRAS.418.1526C, 2015MNRAS.446.3002B, Bonnivard:2016tas, 2016PhRvD..94f3521S, Chiappo:2016xfs, 2016MNRAS.461.2914H, Strigari:2018utn, 2018arXiv180206811P}. 

Second, the $\gamma$-ray emission near each dSph is not modeled to the level of Poisson noise. Errors stem from inaccuracies in diffuse Galactic emission models and the large population of point sources with fluxes too small to be identified. Studies have focused on the latter effect by correlating $\gamma$-ray hotspots with multi-wavelength catalogs~\citep{Carlson:2014nra, Hooper:2015ula, Geringer-Sameth:2018vjd}. However, likelihood-fitting algorithms inexorably link these uncertainties by allowing a combination of point sources and diffuse emission to fit the data over a large region-of-interest (ROI). In Ref.~\citep[][hereafter, L16]{Linden:2016fdd} (see also Appendix~\ref{app:sec:TSdistribution}) we showed that these issues become acute for sources detected at the $\sim$2$\sigma$ level. In this regime, the density of sub-threshold sources approaches the Fermi-LAT angular resolution, and the best-fit model produces a distribution of negative and positive residuals stemming from over- and under-subtraction. 

\begin{figure*}[tbp]
\centering
\includegraphics[width=0.94\textwidth]{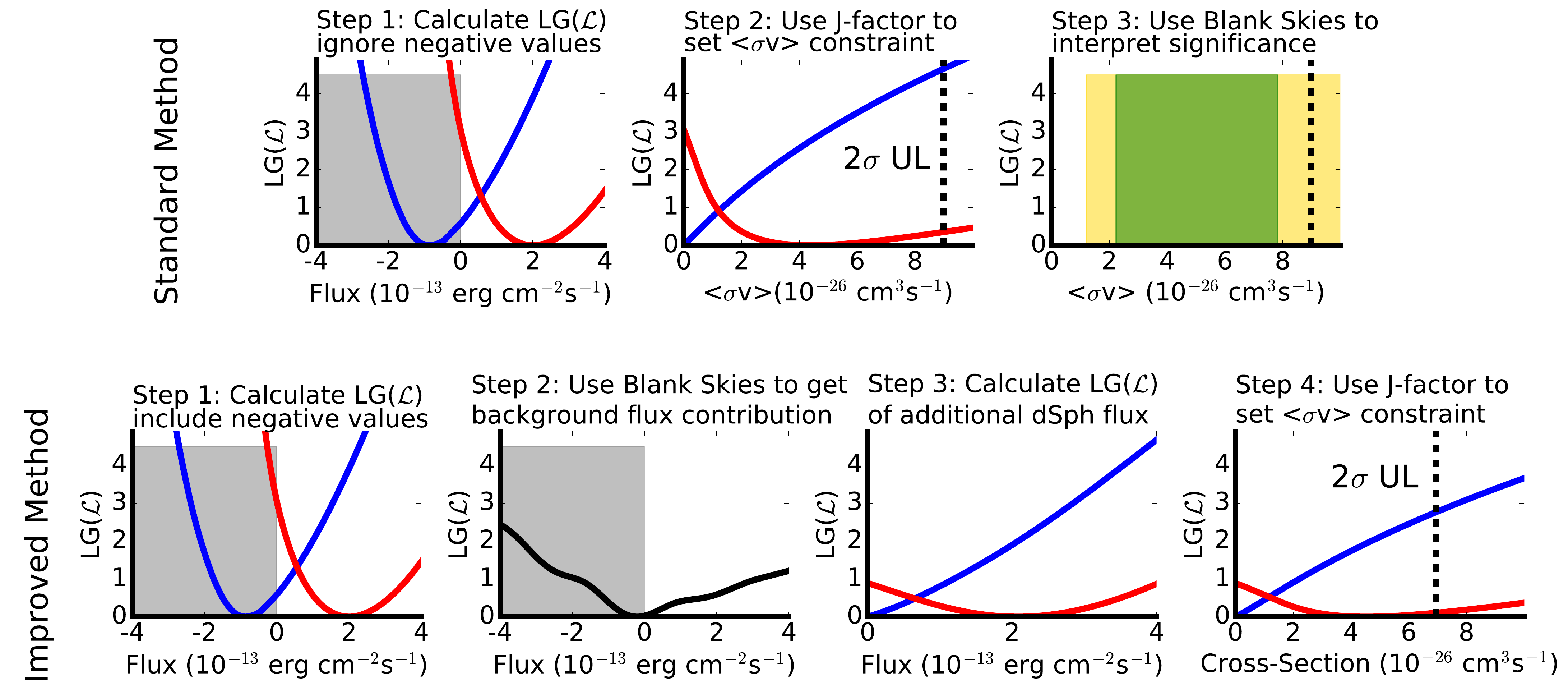}
\vspace{-0.4cm}
\caption{The standard (top) and improved (bottom) joint-likelihood analyses, applied to a two dSph stack of Ursa Major II (blue) and Willman 1 (red). Results are shown for 100~GeV dark matter annihilating to $b\bar{b}$. The key difference is the use of blank sky positions to re-calculate the likelihood profile of each dSph before the joint-likelihood analysis is performed. In this case, the 2$\sigma$ limit is improved by $\sim$30\%. Uncertainty bands can also be calculated in the improved analysis, however the information from blank sky positions is already included to set the limit.}
\label{fig:diagram}
\vspace{-0.3cm}
\end{figure*}

The Fermi-LAT Collaboration's joint-likelihood analysis of dSphs is the quintessential example of an algorithm that operates in this low-significance regime~\citep{Ackermann:2013yva, Ackermann:2015zua, Fermi-LAT:2016uux}. In calculating a 2$\sigma$ exclusion limit evaluated over an ensemble of dSphs, the analysis sets constraints based on low-significance information about each object. While the joint-likelihood algorithm is well-defined when each dSph likelihood profile can be interpreted statistically, the standard algorithm does not account for the possibility that mismodeling induces significant positive or negative residuals coincident with dSph locations. The Fermi-LAT collaboration examines these uncertainties by repeating their analysis on thousands of ``blank sky" locations that do not contain dSphs, finding that the calculated 2$\sigma$ limit can vary by an order of magnitude at 95\% confidence~\citep{Fermi-LAT:2016uux}. This provides an indication for the level of uncertainty produced by diffuse mismodeling, but does not reveal how the true dSph constraint has been affected.

Several data-driven methods have been proposed to constrain the dSph flux without relying on a particular background model~\citep{GeringerSameth:2011iw, Mazziotta:2012ux, GeringerSameth:2012sr, 2015PhRvD..91h3535G, 2015PhRvL.115h1101G, Boddy:2018qur, Calore:2018sdx, Geringer-Sameth:2018vjd, Hoof:2018hyn}. In general, these techniques choose numerous ``blank sky locations" to predict (either through averaging or machine learning) the $\gamma$-ray photon count or flux within a given ROI surrounding each dSph. Comparing the predicted flux with the true dSph flux, they constrain any additional dark matter signal. These analyses benefit because they directly integrate the uncertainty in astrophysical mismodeling into the predicted dSph flux -- accounting for the increased prevalence of low-significance excesses in real data. However, they lose sensitivity in two ways. First, while they utilize $\gamma$-ray data to predict the astrophysical flux near each dSph, they do not model known features (e.g. gas clouds or bright point-sources) that have been observed (and integrated into) astrophysical models. Second, they typically do not take into account the full information provided by the energy-dependent point-spread function in differentiating source events from background.

In this paper, we develop an improved analysis that combines the advantages of the Fermi-LAT joint-likelihood analyses with the accurate treatment of fluctuations provided by data-driven methods. The key conceptual improvement is as follows: while the standard joint-likelihood analysis tests the null-hypothesis that: \emph{the distribution of observed dSph fluxes (weighted by J-factors) is consistent with 0}, our analysis tests the null hypothesis: \emph{the distribution of observed dSph fluxes (weighted by J-factors) is consistent with the distribution of ``fake" point-source fluxes imparted by astrophysical mismodeling}. This integrates the ``blank sky" results directly into the calculated limit. We show that this method produces more resilient detections and constraints on dark matter annihilation, and the results of the new analysis match statistical expectations. Intriguingly, this method can affect the annihilation constraint by a factor of 2 compared to previous work, even when analyzing identical data. We use this model to robustly constrain dark matter annihilation. \newline

\noindent \emph{Joint-Likelihood Analysis ---}  
Our modeling proceeds similarly to L16. Figure~\ref{fig:diagram} illustrates each step in our analysis. For each dark matter model, a probability distribution of point source fluxes induced by astrophysical mismodeling is calculated from an ensemble of blank sky locations as: 
\vspace{-0.2cm}
\begin{equation}
\label{eq:p_bg}
P'_{bg}(\phi_{bg}) = \frac{1}{N}\sum_{i} A_i ~\mathcal{L}_i(\phi_{bg})
\end{equation}
\vspace{-0.3cm}

\noindent where $\mathcal{L}_i$ is the likelihood function of a single source evaluated at flux $\phi_{bg}$, and A$_i$ normalizes the integral of the likelihood function over all fluxes. In Bayesian language, $P'$ can be described as a conjugate prior to the likelihood function. We average over 1000 blank sky positions located at $|b|>$30$^\circ$, and $>$1$^\circ$ from any 3FGL source. The flux can be either positive or negative, because these blank skies account for the impact of mismodeling. The function P$'_{bg}$ is spectrum specific, and is computed for each dark matter mass and final state. 

This definition of P$'_{bg}$ includes contributions from both astrophysical mismodeling and Poisson fluctuations in the photon count. In order to isolate the effect of mismodeling, we calculate the expected flux distribution $\phi_{bg, Poisson}$, expected from statistical fluctuations, and then deconvolute the two effects in order to isolate a function P$_{bg}$, which describes the additional systematic uncertainty. We then use P$_{bg}$ for the remainder of the analysis. Removing the Poisson noise term produces a $\sim$10\% change in our results. In Appendices~\ref{app:sec:TSdistribution} and \ref{app:sec:poisson} we describe this procedure in detail and show both P$'_{bg}$ and P$_{bg}$ for a 100~GeV dark matter particle annihilating to $b\bar{b}$.

\vspace{0.05cm} 
The true $\gamma$-ray flux from a dSph is given by:

\vspace{-0.35cm}
\begin{equation}
\phi_{s,i} = \frac{\sigmav \phi_\chi 10^J}{8\pi m_\chi^2}
\end{equation}
\vspace{-0.25cm}

\noindent where \sigmav~is the annihilation cross-section, J is the logarithmic J-factor, and $\phi_\chi$ is the model-specific energy flux per annihilation. For a given $\sigmav$, the probability that a dSph emits a flux $\phi_{s,i}$ is given by the offset between the calculated J-factor and photometric J-factor:

\vspace{-0.35cm}
\begin{equation}
P(\phi_{s,i}, \sigmav, J_i) = P(J_i) = \frac{1}{\sqrt{2\pi\sigma_i}}~{\rm exp}\left[\frac{-(J-J_{i})^2}{2\sigma_i^2}\right]
\end{equation}
\vspace{-0.2cm}

\begin{figure*}[tbp]
\centering
\includegraphics[width=1.0\textwidth]{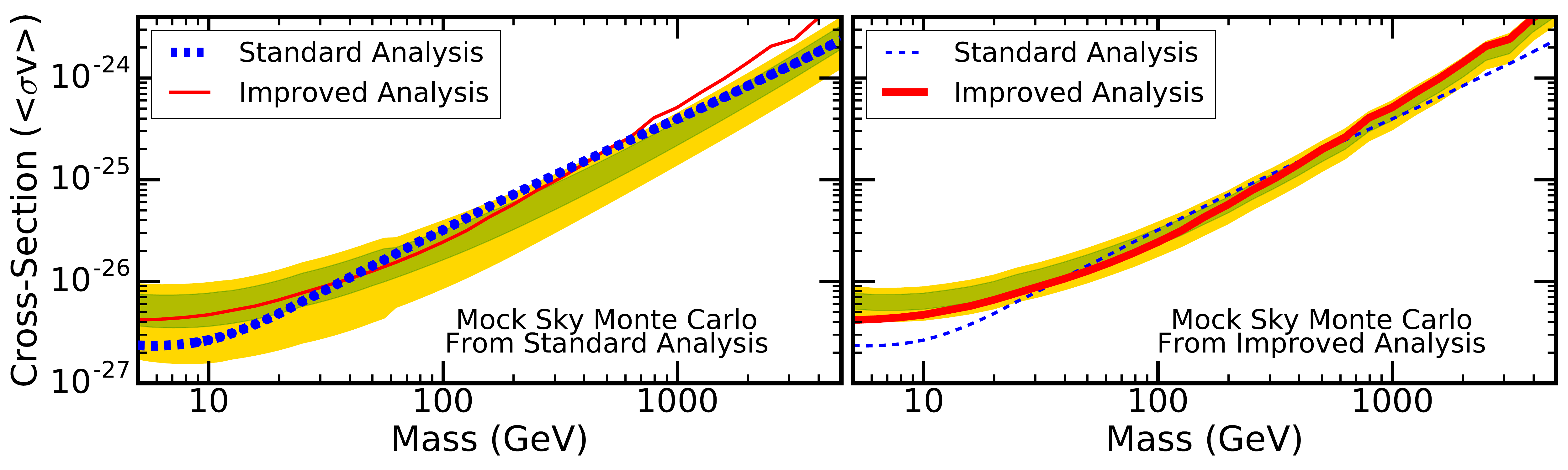}
\vspace{-0.75cm}
\caption{The constraints on dark matter annihilation calculated using the ensemble of 15 dSphs utilized in~\cite{Ackermann:2013yva} for the standard analysis (left) and our improved analysis (right). Results from the opposing analysis are shown with thin lines to aid the eye. We find that the resulting constraints can differ by a factor of two at both low and high masses. In the improved analysis, the uncertainty bands decrease drastically, because deviations produced by astrophysical mismodeling are absorbed by P$_{bg}$. This illustrates the robustness of the approach.}
\label{fig:15systems}
\vspace{-0.35cm}
\end{figure*}

\noindent where J$_{i}$ and $\sigma_i$ are the photometric J-factors and uncertainties for dSph $i$, expressed logarithmically. Given this definition, the joint-likelihood of an ensemble of dSphs is calculated by first integrating out the background  contribution, and then maximizing the resulting likelihood for some value of P($\phi_{s,i}$):

\vspace{-0.3cm}
\begin{multline}
\label{eq:theequation}
P(\sigmav) = \prod_{i} \argmax_{\phi_{s,i}}  \int_{-\infty}^\infty \mathcal{L}(\phi_{s,i} + \phi_{bg}) P_{bg}(\phi_{bg})~ \times \\ \times P_{s, i}\left(\phi_{s,i},\sigmav, J_i, \sigma_i \right) {\rm d \phi_{bg}}
\end{multline}
\vspace{-0.4cm}

\noindent where each dSph flux is a combination of a true source flux ($\phi_{s,i}$) and a flux induced by background mismodeling ($\phi_{bg}$). The source flux is non-negative (it represents a physical source), while the background flux can be negative (it represents the flux induced by mismodeling). The results are weighted by the probabilities (P$_{bg}$ and P$_{s_i}$) that each flux is consistent with the blank sky and dark matter fluxes, respectively. The joint-likelihood is the product of each dSph likelihood. In this model, the probability is maximized at the best-fit value of $\phi_{s,i}$. In Appendix~\ref{app:sec:jfactors} we show results obtained by marginalizing over $\phi_{s,i}$. At their core, these models both integrate the information from blank sky positions into the dSph flux calculation \emph{before} the joint-likelihood is performed. We now demonstrate the advantages of this approach.\newline

\noindent \emph{Fermi-LAT Analysis ---} We analyze 9.5~yr of Fermi-LAT data in a 14$^\circ\times$14$^\circ$ ROI around each dSph. We divide P8R2\_SOURCE\_V6 photons into four bins based on the their angular reconstruction (event types: PSF0--PSF3), and 24 logarithmic energy bins spanning 500~MeV to 500~GeV. We analyze each bin separately and sum the results to obtain the total log-likelihood. We additionally analyze 1000 ``blank sky locations" at latitude $|b|>$30$^\circ$ and separated by  $>$1$^\circ$ from any 3FGL source. We assume that each dSph is represented by a point-source -- rather than a spatially extended profile. Ref.~\citep{Ackermann:2013yva} showed that this only marginally affects the calculated flux -- and we stress that the goal of this letter is to focus on the likelihood algorithms. Thus, we designed this analysis to be similar to Fermi-LAT collaboration studies when possible (a few exceptions are noted in Appendix~\ref{app:sec:fermidata}). 

Unlike previous work, we evaluate both positive and negative fluxes in each energy bin to calculate the dSph likelihood profile ($\Delta$LG($\mathcal{L}$)). In this work, we utilize a two-step fitting algorithm. First, the background diffuse and point source emission is fit to data over the ROI following the standard formalism. We subsequently fit the dSph to the data, allowing only the dSph normalization to float. We have tested alternative methods, and found that they negligibly affect our results when the dSph flux is small, as is the case here. Further analysis details are provided in L16 and Appendix~\ref{app:sec:fermidata}.

\noindent \emph{Results ---} We reproduce two Fermi collaboration studies, comparing the standard joint-likelihood analysis with our method. We first focus on the 15 dSphs examined by~\citep{Ackermann:2013yva, Ackermann:2015zua} and then the 45 dSphs studied by~\citep{Fermi-LAT:2016uux}. We utilize the reported locations, distances and J-factors from previous work, but use 9.5~yr of data. We make these choices to isolate the effect of the joint-likelihood analysis on the annihilation constraint. \emph{We stress that any change in the constraint stems directly from the analysis method. The datasets are identical, thus the results cannot be considered to be ``statistically consistent."}

\begin{figure*}[tbp]
\centering
\includegraphics[width=1.0\textwidth]{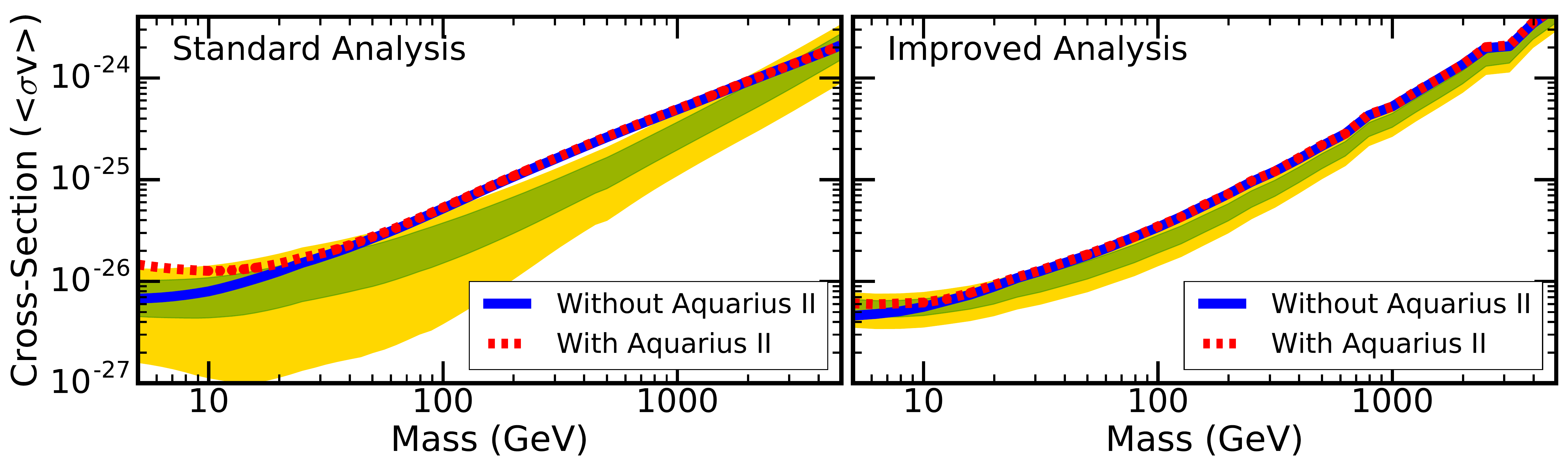}
\vspace{-0.6cm}
\caption{The constraint on dark matter annihilation from the ensemble of 45 dSphs utilized in~\citep{Fermi-LAT:2016uux}. Constraints are shown without (blue solid) or with (red dashed) the addition of the Aquarius II dwarf, which has a significant excess at low $\gamma$-ray energies that is unlikely to be caused by dark matter annihilation. This inclusion of this system significantly affects the standard analysis (weakening the combined limit from 45 dSphs by more than a factor of 2), while our improved analysis accurately treats the tension between the high flux and the low J-factor of Aquarius II. }
\label{fig:45systems}
\end{figure*}

In Figure~\ref{fig:15systems}, we constrain the annihilation cross-section using the standard (left) and improved (right) methods, and show 1$\sigma$ and 2$\sigma$ error bands derived from the blank sky analyses. We note three effects. First, the uncertainty in the expected limit decreases significantly in the improved analysis, demonstrating the resilience of the method. In previous work, the limits varied based on the diffuse mismodeling near individual dSphs. Our model accounts for these fluctuations, decreasing the uncertainty from outlier dSphs.

Second, while the limits are similar between $\sim$50--500~GeV, we find weaker constraints (by a factor of 2) at low masses. This is primarily due to the oversubtraction of low-energy \mbox{$\gamma$-rays} near Segue I, which is the highest J-factor dSph in the sample. The best-fit flux for 5~GeV dark matter in Segue I is \mbox{-7$\times$10$^{-14}$~erg~cm$^{-2}$~s$^{-1}$}. Because the standard model does not allow mismodeling to induce this flux, it adopts a likelihood function that is skewed towards 0. This forces the low-mass constraint to fall nearly 2$\sigma$ below expectations. In the improved analysis, the negative flux from Segue I is absorbed into $\phi_{bg}$. Our analysis still obtains a strong limit from Segue I, because $\phi_{s,i}$ cannot be too big without $\phi_{bg}$ becoming unreasonably negative. Indeed the improved constraint is still $\sim$2$\sigma$ below expectations, but the amplitude of the shift is smaller.

Third, the high-mass limits are also a factor of $\sim$2 weaker in our analysis. This is not a fluctuation, but a robust feature (note that the expected error bands predict weaker constraints). The energy flux from TeV dark matter peaks at $\sim$100~GeV, where few photons are present. It is likely that 0 photons are observed near the dSph, while the background model predicts $\mathcal{O}(0.1)$ events.  Thus, the average dSph has a slightly negative best-fit flux. In the standard analysis, this significantly strengthens the combined limit. However, the improved analysis is normalized to blank-sky positions that also skew towards negative fluxes, correcting for this effect.

In Figure~\ref{fig:45systems} (blue), we show constraints from the 45 dSphs analyzed in~\citep{Fermi-LAT:2016uux}. In this case, we find that our analysis produces stronger constraints (by $\sim$25\%) below $\sim$500~GeV. The stronger limits derive from the small excesses found in several dSphs (e.g. Reticulum II), which cause the standard analysis to prefer the annihilation of 150~GeV dark matter to $b\bar{b}$ at TS~=~8.9 (similar to Ref.~\citep{Fermi-LAT:2016uux}). Nominally, this would translate to a 3$\sigma$ detection of dark matter. However, an excess of this significance is relatively common in analyses of blank-sky locations, decreasing our confidence in this result. The improved analysis also finds a slight excess at 150~GeV (because these dSphs have real upward fluctuations). However, after accounting for the probability that background mismodeling induces similar upward fluctuations, it finds the significance of this excess to be only TS=0.4.

\begin{figure*}[tbp]
\centering
\includegraphics[width=0.95\textwidth]{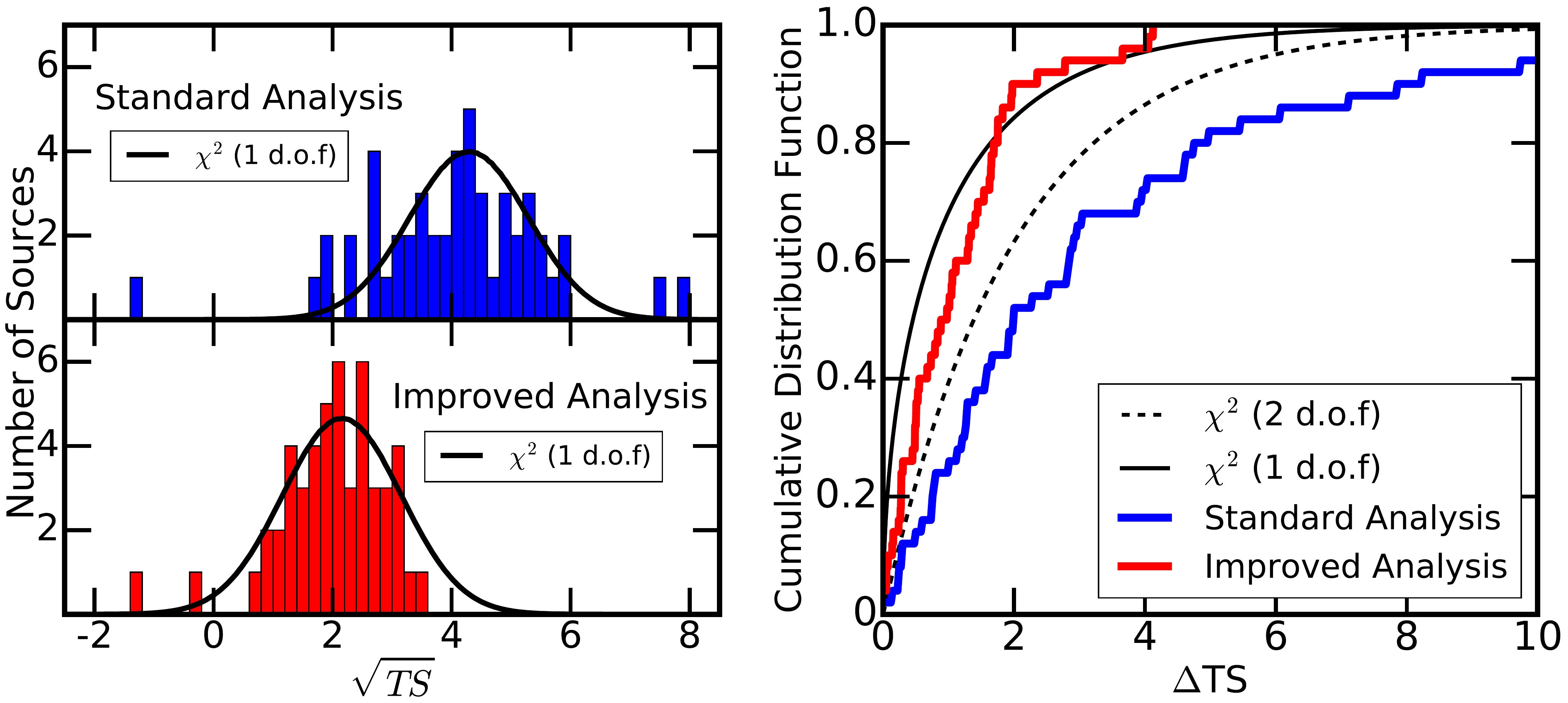}
\vspace{-0.4cm}
\caption{ (Left) - The recovered statistical significance from 50 trials including an injected 100~GeV dark matter signal with cross-section of 3$\times$10$^{-26}$~cm$^{3}$s$^{-1}$. In the default analysis (top left), the statistical significance includes several significant upward and downward fluctuations that would not be expected from a $\chi^2$ distribution surrounding the average significance (such a distribution is rejected at 5.1$\sigma$). The improved analysis (bottom left) closely follows a normal distribution. (Right) - The cumulative distribution function of the $\Delta$TS between the injected dark matter mass and cross-section and the recovered best-fit dark matter mass and cross-section in the standard (blue) and improved (red) analyses. Results are compared $\chi^2$-distributions with 1 and 2 d.o.f. (as the mass is only scanned coarsely). The improved analysis follows this expectation, while the standard analysis prefers an incorrect mass and cross-section at a significance TS$>$4 in $>$30\% of cases.}
\label{fig:injection}
\end{figure*}

In Figure~\ref{fig:45systems} (red-dashed), we show a scenario that challenges the standard analysis, induced by adding the \mbox{Aquarius II} dSph into the stack. Aquarius II has a bright, soft-spectrum $\gamma$-ray excess. Modeling this excess with 5~GeV dark matter (to $\bar{b}b$) improves the log-likelihood by 10.7, statistically representing a 4.6$\sigma$ detection of dark matter. However, Aquarius II is not expected to be bright, its logarithmic J-factor is only 18.27$\pm$0.62~\citep{Pace:2018tin}. The standard algorithm resolves this discrepancy by treating the J-factor measurement and likelihood function equivalently, assuming  that the excess is statistical in nature. Thus, the algorithm increases the Aquarius II J-factor to 19.96, paying a log-likelihood cost of 6.9. This produces a TS=14 detection of 5~GeV dark matter in the joint-likelihood analysis and weakens the limit by a factor of 2. 

The improved algorithm includes the possibility that the Aquarius II flux is produced by $\phi_{bg}$. Blank sky analyses indicate that there is a $\sim$0.7\% chance that a background fluctuation would induce an excess compatible with 5~GeV dark matter at this significance. Thus, the Aquarius II result is significant at $\lesssim$2.4$\sigma$. Combined with its low J-factor, this prevents the improved algorithm from overreacting to the excess. This test illuminates two lessons. First, the improved analysis allows the statistical uncertainties in each dSph J-factor to be directly compared to the significance of any $\gamma$-ray excess. Second, the analysis can accurately treat outliers in the Fermi-LAT data, whereas the standard method would require manual intervention to remove Aquarius II. \newline

\noindent \emph{Mock Signal ---} Thus far, one should worry that our analysis can only constrain dark matter, and may misinterpret a true signal as a fluctuation in $\phi_{bg}$. Here, we show that the analysis accurately recovers signals that are injected into $\gamma$-ray data. We produce 50 simulations of 15 blank sky locations at latitude $|b|>$30$^\circ$ lying $>$1$^\circ$ from any 3FGL source. We assign each blank sky to be one of the 15 dSphs analyzed in~\citep{Ackermann:2013yva} and use {\tt gtobssim} to inject a signal corresponding to the best-fit J-factor of each dSph and a 100~GeV dark matter model that annihilates to $b\bar{b}$ with a cross-section of 3$\times$10$^{-26}$~cm$^{3}$s$^{-1}$. This lies near the 2$\sigma$ constraint shown in Figure~\ref{fig:15systems} for both analyses. Thus, we expect a typical detection at $\sim$2$\sigma$.  

In Figure~\ref{fig:injection}, we show that standard analyses produce results that vary from a -1.4$\sigma$ exclusion of the signal to a 7.9$\sigma$ detection. The hypothesis that this dispersion is consistent with Gaussian fluctuations around the mean signal significance is rejected at 5.1$\sigma$ (still 3.3$\sigma$ even when the worst outlier is removed). Scanning the 2D mass/cross-section space, we find that the incorrect mass and cross-section are preferred at TS$>$4 in $>$30\% of cases. While blank skies analyses could characterize this uncertainty, the standard method would have difficulty indicating the correct parameters. 

Our analysis produces a distribution of significances following a $\chi^2$ distribution centered near the predicted value of $\sim$2$\sigma$. While the distribution of $\sqrt{TS}$ is slightly more clustered than expected, this is significant at only 0.9$\sigma$. In the two-dimensional reconstruction, the variance between the injected and recovered parameters also follows a $\chi^2$ distribution. We note that the dark matter mass and the $\gamma$-ray energy are scanned coarsely, and the mass and cross-section parameters are not linearly independent. We thus expect to observe a $\chi^2$ distribution with between 1---2 d.o.f. We conclude that our method is sensitive to dark matter signals, and moreover, that the significance of any discovered excess can be interpreted using standard statistical methods. \newline

\noindent \emph{Dark Matter Constraints ---} We now evaluate the annihilation constraints shown in Figures~\ref{fig:15systems}~and~\ref{fig:45systems}, finding that our analysis rules out (at 95\% confidence) dark matter masses below 120~GeV and 87~GeV annihilating to $\bar{b}b$ at a cross-section of 3$\times$10$^{-26}$~cm$^{-3}$s$^{-1}$. The constraints do weaken after the inclusion of 30 additional dSphs (and some J-factor changes between~\citep{Ackermann:2013yva}~and~\citep{Fermi-LAT:2016uux}). This is a real effect, based on the low-significance detection of excesses in Reticulum II and a handful of other dSphs. Our results produce maximum TS values of $\sim$0.04 at a mass of 5~TeV with 15 dSphs, and 1.06 at a mass of $\sim$31~GeV with 45 dSphs. These TS values can be interpreted as true statistical preferences for dark matter, though at a very low significance. We argue that these results, utilizing 9.5~yr of data and an improved statistical method, set robust constraints on dark matter annihilation.\newline


\noindent \emph{Conclusion ---} In this \emph{paper} we introduced a simple, but significantly more powerful, method to evaluate the constraints on annihilating dark matter through the joint-likelihood analysis of dwarf spheroidal galaxies. We have demonstrated several advantages to this technique using real Fermi-LAT data: (1) the technique is resilient to systematic uncertainties in Fermi-LAT background modeling and does not produce overly strong limits or unreasonably strong excesses due to diffuse over- or under-subtraction, (2) the technique allows the statistical significance of excesses to be directly compared to the statistical significance of the J-factor measurement, without biasing results towards the Fermi-LAT measurements that are not represented by Poisson statistics, (3) the technique accurately measures the uncertainty in both the cross-section and mass of an injected dark matter signal, and does not produce high precision (but low accuracy) measurements.

We note that this method integrates seamlessly with existing analyses conducted by the Fermi-LAT collaboration, only requiring the use of Eq.~\ref{eq:theequation} to evaluate the joint-likelihood function. We thus recommend its integration in future dSph analyses.  Moreover, as noted in L16, the result can be trivially adapted to any search for low-significance excesses in Fermi-LAT data, potentially improving our sensitivity to new phenomena on the threshold of detection. In Appendix~\ref{app:sec:technicaldetails}, we discuss several technical considerations required in the implementation of these analysis techniques.

Finally, the result of this analysis provides of the most robust constraints on annihilating dark matter. Using 9.5~yr of Fermi-LAT data, we rule out the thermal annihilation cross-section up to masses of $\sim$87~GeV. However, several simplifications have been made (due to our focus on statistical techniques), which will be improved in an upcoming publication. Most notably, we have not utilized the most up to date photometric J-factors produced by numerous groups, instead choosing values consistent with previous studies. Second, we have modeled dSphs as point sources, rather than carefully considering their spatial extension. Third, we have not used the very recently released 4FGL catalog, which appeared subsequent to the computations shown in this paper. The inclusion of these improvements into the current framework is straightforward, and will be considered in an upcoming publication.

\section*{Acknowledgements}
I would like to thank John Beacom, Francesca Calore, Jan Conrad, Alex Drlica-Wagner, Dan Hooper, Jason Kumar, Simona Murgia, and Annika Peter for providing feedback that significantly improved the quality of this paper. I especially would like to thank Alex Geringer-Sameth for a number of valuable comments that provided significant insight into the statistical formalism described here. Portions of this project were completed in part at the Aspen Center for Physics, which is supported by National Science Foundation grant PHY-1607611. This project also benefited from significant technical support from the Ohio Supercomputing Center.

\bibliography{dwarf_statistics}

\newpage

\appendix

\section{The Density of Sub-Threshold Sources as a Function of TS}
\label{app:sec:TSdistribution}

In this section, we discuss the distribution of positive and negative residuals in the Fermi-LAT data, and their influence on the TS distribution of $\gamma$-ray point sources. We also refer the reader to L16, which provides additional information regarding our statistical methods. The influence of unidentified point-sources is significant in Fermi-LAT analyses due a combination of its high sensitivity ($\sim$10$^{-12}$~erg~cm$^{-2}$~s$^{-1}$) and modest angular resolution ($\sim$1$^\circ$ for individual 1~GeV photons). This combination results in the detection of a handful of sources observed at the $\sim$2$\sigma$ level within every angular resolution element -- and implies that unidentified sources contribute to the photon count everywhere in the $\gamma$-ray sky.

\begin{figure}[tbp]
\centering
\includegraphics[width=.48\textwidth]{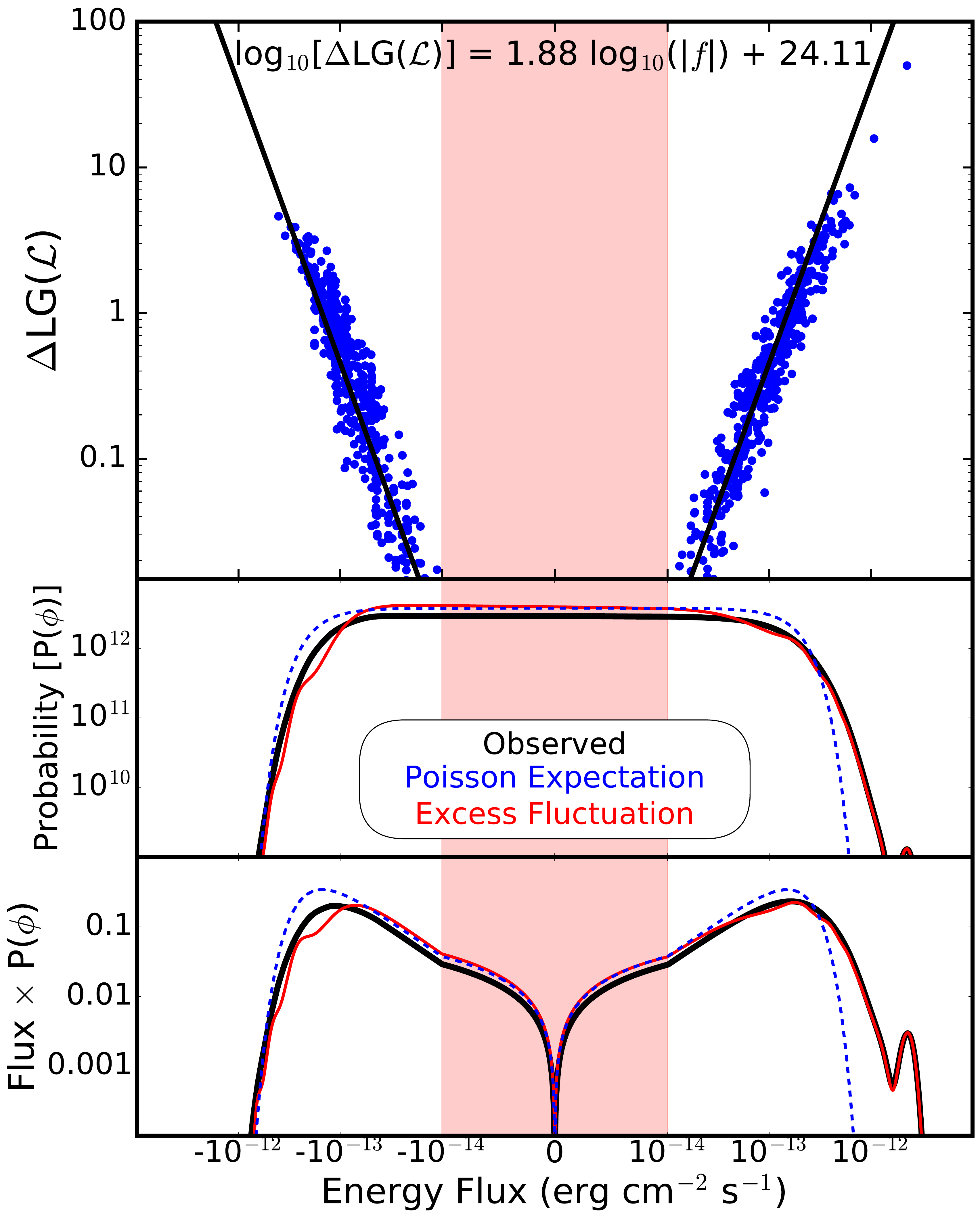}
\caption{ (Top) The relationship between the point source flux and the improvement of the log-likelihood fit to the data. Points are taken from 1000 blank sky positions, assuming a $\gamma$-ray spectrum produced by the annihilation of 100~GeV dark matter to $b\bar{b}$. The improvement in the log-likelihood and the best-fit flux are tightly correlated, as would be expected given that the $\Delta$LG($\mathcal{L})$ is based on the photon count in each energy and angular bin. Importantly, this relationship holds for $\gamma$-ray sources with both positive and negative fluxes, illustrating the importance of accurately quantifying negative fluctuations produced by modeling uncertainties. (Middle) The relative probability that a blank sky position has a particular point-source flux for the same dark matter model (black solid). The distribution of positive and negative fluxes is nearly symmetric up to a flux of nearly $\sim \pm$2$\times$10$^{-13}$ erg~cm$^{-2}$s$^{-1}$, which roughly translates to $\Delta$LG($\mathcal{L})$~=~2. The flux is broken down into the expected distribution from Poisson fluctuations (blue dotted), and the residual which is due to the mismodelling of the background (red solid). (Bottom) Same as the Middle, but multiplied by the $\gamma$-ray flux to show the regions where sources are most likely to be located. A large, significant residual is seen, especially at high positive fluxes.} 
\label{fig:residuals}
\end{figure}

To quantify these statements, we note that there is a strong correlation between the calculated flux of a $\gamma$-ray source and the TS of the same source. In regions that are background dominated (the number of source photons is much smaller than the number of background photons), one would expect the source TS to scale as the square of the number of photons attributed to the source. This is due to the fact that (so long as there are many photons) the result is normally distributed:

\begin{equation}
\sigma_{s} = \sqrt{TS} = \frac{(n_s + n_{bg} - n_{bg})}{(\sqrt{n_{bg} + n_{s}})} \approx \frac{n_s}{\sqrt{n_{bg}}}
\end{equation}

\noindent where n$_{bg}$ is the number of events expected (and observed) from the background, while n$_s$ is the number of events produced by the source. In Figure~\ref{fig:residuals}, we show the results of a direct calculation of this correlation over our ensemble of blank sky positions. We find that for a spectrum fitting a dark matter mass of 100~GeV annihilation to $b\bar{b}$, the best-fit powerlaw spectrum is TS~$\propto \phi^{1.88}$. Intriguingly, this correlation holds for both positive and negative values of the $\gamma$-ray flux. 

To determine the density of sources that contribute such a flux to the $\gamma$-ray data, we note that the 3FGL catalog includes 1307 sources detected at $|b|>$30$^\circ$ with a TS$>$25. Furthermore, the flux-distribution of these sources (at fluxes far from the Fermi-LAT detection threshold) is approximately dN/df~$\propto$~f$^{-2}$~\citep{Acero:2015hja}. The majority of these sources are extragalactic AGN, which have power-law flux distributions that continue to very low fluxes~\citep{2012ApJ...751..108A}. Thus, the observation of 1307 sources in our ROI, along with the TS-flux correlation identified above, implies the existence of $\sim$5000 sources contributing at the level TS$>$2, for detections using only the 4 years of data included in the 3FGL catalog. Moving to 9~years of data (noting that the TS of a true source should scale linearly with exposure), indicates that the density of TS$\sim$5 sources is approximately $\sim$0.25~deg$^{-2}$, similar to the $\sim$1$^\circ$ PSF obtained when only a handful of $\gamma$-rays are observed.

Because there is $\mathcal{O}$(1) source that contributes at TS$\sim$5 in each PSF element scanned by the Fermi-LAT, and furthermore, because this source is poorly localized, these sub-threshold sources produce $\sim$2$\sigma$ excesses at a much higher rate than expected from Poisson statistics. Moreover, regions with $\sim$1 such source will have positive residuals, while regions with 0 sources will have negative residuals. Figure~\ref{fig:residuals} shows this distribution, finding a nearly 50\% chance that an analysis of a random sky position has its log-likelihood improved when a negative point source is added to the analysis.  This is due to a combination of both modeling errors, as well as Poisson fluctuations in the photon count.

In Figure~\ref{fig:residuals} (middle), we show the resulting distribution of P$'_{bg}$ for a model with a 100~GeV dark matter particle annihilation to $b\bar{b}$ in our analysis. This shows that there is a relatively equal probability that the flux has either positive or negative values up until a TS of approximately 6 (LG({$\mathcal{L}$)$\sim$3). At higher-values, there is a positive excess, owing to brighter point sources that contribute at higher-flux values. Furthermore, we note that for fluxes smaller than $\sim$10$^{-13}$~erg~cm$^{-2}$s$^{-1}$, the probability distribution is roughly flat, and symmetric for positive and negative values. Figure~\ref{fig:residuals} (bottom), we show the same flux distribution, but multiplied by the flux in order to illustrate the regions of parameter space where sources are most likely to appear. We note that, given a probability distribution that is flat over many decades in flux, random positions are highly likely to have fluxes near the value of $\sim$10$^{-13}$~erg~cm$^{-2}$s$^{-1}$, compared to fluxes with much smaller absolute values. In the next section we describe the deconvolution of our observed data (black solid lines) into components relating to statistical and systematic errors. 

\vspace{-0.3cm}
\section{Separation of Poissonian and Systematic Errors}
\label{app:sec:poisson}
\vspace{-0.2cm}

Because the blank sky analysis operates on real Fermi-LAT data, the likelihoods that enter into Equation~\ref{eq:p_bg} include both systematic errors (which we would like to isolate and account for in P$_{bg}$) and Poisson fluctuations (which should not be included in the calculation of P$_{bg}$, because we want the calculated fluxes of dSphs to include statistical fluctuations). To deconvolve these two terms, we note that the Poissonian error nearly follows a Gaussian distribution, and we calculate the distribution of fluxes induces by Poissonian fluctuations using the correspondence between the $\gamma$-ray flux and the log-likelihood fit to the $\gamma$-ray data (Figure~\ref{fig:residuals} (top)). In this paper, we build a model for the Poissonian fluctuations by calculating the average Poisson error in all blank sky locations. We note that this technique could be slightly improved by instead calculating the individual Poisson errors in each blank sky location (since the photon counts differ slightly near each blank sky position). However, because we use blank sky locations that lie at high Galactic latitude and far from any point sources, the background fluxes for all sources are similar. 

In theory, many deconvolution algorithms can be used to separate these two components -- but preventing numerical issues can be tricky (due to the large dynamic range in P$_{bg}$).  In this paper, we utilize forward modeling. We first guess at the systematic distribution P$_{bg}$ and then convolve this distribution with the Poissonian noise term to fit the measured value of P$'_{bg}$. We iteratively improve this fit and approach a ``good" solution. After each iteration we use a high-pass filter to remove high-frequency noise. We obtain solutions that fit to within 5\% (usually within 1\%) over the range where the P$_{bg}$ is more than 10$^{-5}$ of its maximum value. This fit could be improved with additional iterations. The result for 100~GeV dark matter is shown in the middle and lower panels of Figure~\ref{fig:residuals}. We find that the amplitude of the excess is much larger for positive flux values -- representing the appearance of real sources that produce fluctuations above the expected level of Poisson noise. However, there are important contributions to P$_{bg}$ at negative fluxes as well. In particular, we note that the similarity between the Poisson expectation of P$'_{bg}$ and the observed value of P$'_{bg}$ is incidental. If only positive systematic errors existed in the Fermi-LAT data, we would expect P$'_{bg}$ to have fewer negative fluctuations than expected by our Poisson model.

\section{Fermi Data Analysis}
\label{app:sec:fermidata}
\vspace{-0.2cm}

In this section, we provide further details regarding the Fermi-LAT data analysis. We examine events recorded between MET times 239557417--548377407 that pass the P8R2\_SOURCE\_V6 event class, were recorded at a zenith angle below 90$^\circ$, and an energy between 500~MeV--500~GeV. These events are divided into four classes based on the reliability of their angular reconstruction (evtype=4,8,16,32), and all analyses proceed independently for each PSF class. The source likelihood function is calculated from the individual likelihoods of each PSF class. Each set of events is binned into 140x140 pixels with a bin size of 0.1$^\circ$, centered on the position of a dSph (or blank sky) source. Events are binned into 24 logarithmically equal energy bins. The ROI is slightly larger than that chosen for the Fermi-LAT analysis, which utilized a 100x100 grid with identical 0.1$^\circ$ pixels. This difference potentially explains the different sensitivity of our low-energy Monte Carlo constraints compared to the Fermi-LAT results~\citep{Ackermann:2015zua}. However, it is also possible that differences in the dark matter spectral models at low-masses affect the results. The instrumental exposure is calculated in 0.5$^\circ$ increments across the spatial map in the same 24 energy bins utilized for the analysis. The exposure is calculated using the EDGES option, which calculates the instrumental exposure at both edges of the energy bin and interpolates the result.

We first calculate the likelihood fit of a background model in each sky region. This fit includes all 3FGL sources, as well as the gll\_iem\_v06.fits diffuse model and the iso\_P8R2\_SOURCE\_v6\_PSFX\_v06.txt isotropic background model, where X represents the PSF class of each analysis. We calculate the likelihood function (and flux) of each source independently in each energy bin. As such, we allow only the total normalization of each source to vary, with the spectrum of each source fixed to the best-fit 3FGL properties. Because we are using small energy bins, this only marginally affects our analysis. In this stage, we do not include a source at the tentative position of the dSph or blank sky position. The source model is produced with {\tt gtsrcmaps} and the likelihood fit evaluated with {\tt gtlike}, with energy dispersion disabled.

\begin{figure*}[tbp]
\centering
\includegraphics[width=1.0\textwidth]{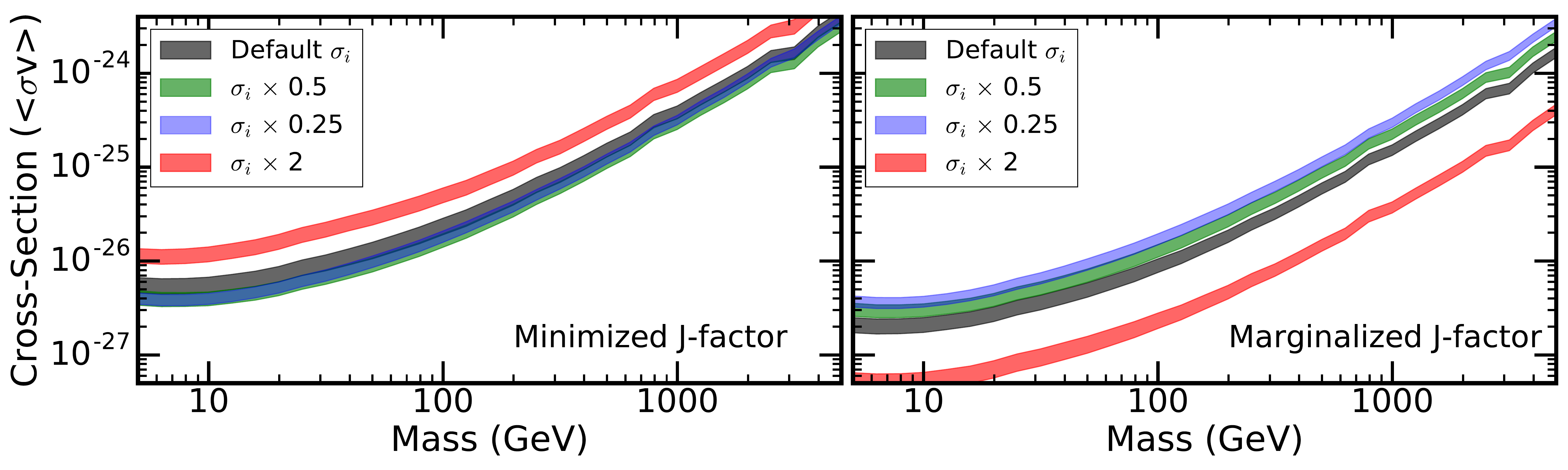}
\vspace{-0.55cm}
\caption{The 1$\sigma$ uncertainties in the expected constraint on dark matter annihilation obtained from the stacked analyses of 45 dSphs described in~\citep{Fermi-LAT:2016uux}. Results are shown using two different methods for including J-factor uncertainties into the joint-likelihood analysis, including the analysis described by Equation~\ref{eq:theequation} (left), and an analysis that marginalizes over the J-factor uncertainty following Equation~\ref{eq:theequationmarginal} (right). In both cases, we show the default limits (black), as well as limits from scenarios where we multiply the J-factor uncertainty in each individual dSph by 2 (red), or divide the J-factor uncertainty by 2 (green) and 4 (blue). In our default analysis, the limit behaves as expected, producing stronger constraints as the uncertainty on the J-factor of individual dSphs decreases. In the marginalized analysis, the limit demonstrates perverse behavior, becoming stronger as the J-factor uncertainty increases. This is due to the very strong constraints on large upward fluctuations in the dSph J-factor provided by $\gamma$-ray analyses.}
\label{fig:marginalcomparison}
\end{figure*}

We then evaluate the expected flux from a point source added at each dSph or blank sky position. We utilize {\tt gtmodel} to generate the expected number of counts as a function of position and energy for a given source normalization, and then calculate the improvement to the total log-likelihood as a function of the source flux in each energy bin. During this stage the background model is fixed. We allow the flux of the putative point source to be either positive or negative, representing the additional contribution to the point-source flux from background mismodeling. 

We note that Poisson likelihoods cannot be used when the total model prediction (including both the dSph and the background model) is negative. This scenario is very rare, and present only in the highest energy bins, which contribute negligibly to the total LG($\mathcal{L}$). However, to prevent numerical issues in likelihood scans, we treat this scenario as follows. If the model predicts a negative number of counts in a specific angular and energy bin, we first check the number of observed counts in the data. If the number of observed counts is non-zero, we assign an infinite likelihood due to continuity (the log-likelihood of predicting 0 counts would also be infinite). If the number of observed counts is 0, we allow the predicted flux to be negative, and assign a log-likelihood based on the absolute value of the predicted model counts. This maintains a local minimum at the best-fit prediction of zero. We stress that this scenario rarely arises near the best-fit values of the flux, and is important only for the stability of algorithms.

\section{Marginalizing Over J-Factor Uncertainties}
\label{app:sec:jfactors}

In our default analysis, we determine the flux from each individual dSph by integrating over the possible background contributions to the total flux to determine the best-fit value of the source flux $\phi_{s,i}$. However, we then minimize to find the best-fit value of $\phi_{s,i}$. Because we are testing a specific value of $\sigmav$ in a specific dSph, this is equivalent to finding a best-fit J-factor in each dSph (where best-fit includes the penalty from choosing a J-factor that deviates from the observed value). In this sense, the method is pseudo-Bayesian, as it builds a prior for the systematic errors, but minimizes the fit for some values of the dSph J-factors. Alternatively, we could institute a more Bayesian approach, where we marginalize our results over all possible J-factors that could contribute to the fit, as follows:

\vspace{-0.2cm}
\begin{multline}
\label{eq:theequationmarginal}
P(\sigmav) = \prod_{i} \int_{-\infty}^\infty \int_0^\infty \mathcal{L}(\phi_{s,i} + \phi_{bg}) P_{bg}(\phi_{bg})~ \times \\ \times P_{s, i}\left(\phi_{s,i},\sigmav, J_i, \sigma_i \right) \; {\rm d\phi_{s,i}}\, {\rm d \phi_{bg}}
\end{multline}
\vspace{-0.3cm}

\noindent where the likelihood and probability distributions are defined as in the main text, while the integral over $\phi_{s,i}$ is assumed to have a logarithmically-flat prior. The integral over $\phi_{s,i}$ includes only non-negative values, as the emission represents a true source that must be non-negative. This is identical to a constraint that the dSph J-factor must be non-negative.

This technique potentially improves the sensitivity of dark matter searches, as it includes additional information regarding the distribution of J-factor uncertainties. For example, while Equation~\ref{eq:theequation} will not impose any costs to a scenario where the J-factor of each dSph falls exactly at the predicted value, Equation~\ref{eq:theequationmarginal} imposes likelihood costs for any scenario where a portion of the J-factor parameter space that is consistent with optical velocity dispersion measurements begins to be ruled out. On the other hand, this means that Equation~\ref{eq:theequationmarginal} assumes detailed knowledge of not only the best-fit J-factors, but also of the uncertainty in each J-factor calculation. Moreover, Equation~\ref{eq:theequationmarginal} assumes that the uncertainties in each dSph J-factor are uncorrelated, which may or may not hold depending on the analyses utilized in the J-factor calculations.

In Figure~\ref{fig:marginalcomparison}, we show the constraints from each method, applied to the Monte Carlo analyses of the 45 dSphs analyzed in~\citep{Fermi-LAT:2016uux} (the differences in the cross-section constraint in the case of 15 dSphs are significantly smaller, with an amplitude of $\sim$20\%). We find that utilizing Equation~\ref{eq:theequationmarginal} produces a constraint on the annihilation cross-section that is nearly a factor of three stronger than our standard analysis (given by Equation~\ref{eq:theequation}). This is similar to the recent result of~\cite{Hoof:2018hyn}, who found that Bayesian analyses of dSphs produced significantly stronger limits. This result is due to the fact that the J-factor uncertainties are logarithmic, and significant upward fluctuations in the J-factor are strongly constrained by the $\gamma$-ray analysis. Intriguingly, such a method would rule out the thermal annihilation cross-section for dark matter annihilation to $b\bar{b}$ final states up to masses of more than 200~GeV. 

However, marginalizing over the J-factor uncertainties introduces peculiar statistical properties into the calculated limit. It is clear that in the case where the J-factor uncertainty goes to 0, the two limits must agree. In Figure~\ref{fig:marginalcomparison}, we show the calculated dark matter annihilation limit in our default analysis behaves as expected -- as the J-factor uncertainty decreases, the limits become stronger as downward fluctuations of the dSph J-factors become inconsistent with the data. However, in the marginalized case, the limits become weaker as the uncertainty in the J-factor calculation decreases. This is due to the fact that the strength of the marginalized constraint derives primarily from the strong exclusion of upward fluctuations in the J-factor of numerous dSphs. As the fraction of the total J-factor probability that includes large upward fluctuations increases, the constraint on the dark matter annihilation cross-section becomes stronger.

The legitimacy of each interpretation is open for debate. It appears reasonable that the marginalized model should be used in scenarios where the uncertainty in each individual dSph is understood and is statistically independent from the uncertainty in other dSphs. In this case, the stacked analysis of 45 dSphs should rule out scenarios where \emph{none} of the 45 dSphs in the analysis were consistent with upward fluctuations in the observed J-factor. However, these assumptions are unlikely to hold -- as the calculation of the J-factor depends on the careful evaluation of a number of systematic uncertainties that are likely to be correlated between individual dSphs. Thus, we find the performance of our default J-factor model to provide a more accurate representation of our best constraints on the dark matter annihilation cross-section.

\section{Further Technical Details}
\label{app:sec:technicaldetails}

In this section, we communicate several more technical details from our analysis method. This includes detailing several simplifications utilized to conserve computational time during this study, as well as several caveats and warnings to researchers who may wish to utilize this technique. 

We first note several simplifications in this analysis. In all cases, we believe that these negligibly affect our results. First, following the choice made by the Fermi-LAT collaboration, we have only investigated blank sky positions with latitude $|b|>$30$^\circ$, despite the fact that the population of 45 dSphs analyzed in~\citep{Fermi-LAT:2016uux} includes several objects that are closer to the Galactic plane. We have made this choice in order to produce an ``expected constraint" plot which is similar to the Fermi-LAT results, but note that the correct choice of blank sky locations is somewhat more important in our analysis --- as the blank sky locations are directly utilized in order to calculate the dark matter annihilation limit. In future work, we will select an ensemble of blank sky locations which is more closely matched to the analyzed dSph population. 

Second, in generating the expected dark matter limits, we have utilized the same blank sky population that was employed to produce the distribution P$_{bg}$ in the joint-likelihood analysis. This technically removes some statistical independence from the blank sky locations used to produce the expected dark matter constraints, since each dwarf contributes at the 0.1\% level to the P$_{bg}$ calculation. Thus, this is expected to narrow the range of uncertainties in our calculated limits at the order of a few tenths of a percent (since a handful of dSphs dominate the result in any dwarf stacking analysis). This was done to save computational time, and will be tested in detail in future analyses. 

Third, in our analysis of the expected limits for 45 dSphs, we have doubled the number of Monte Carlo trials by re-calculating the expected limit when each blank sky location is re-assigned from representing the highest J-factor dSph to representing the lowest J-factor dSph (and vice versa). This decreases the statistical independence of our results by a small amount --- because the 22 dSphs with the lowest J-factors contribute negligibly to the limit. This was done to double the number of trials in the dSph Monte Carlo, since more than 22 trials were necessary to generate reasonable 1$\sigma$ and 2$\sigma$ errors. We used an entirely independent set of Fermi-LAT sky positions to calculate the expected signal when dark matter is injected into the data, keeping that analysis independent.

We additionally note several technical details for those wishing to replicate these results. First, the utilization of negative values for the point source flux can produce mathematical issues whenever the expected counts from the entire model (source + astrophysical background) becomes negative, as the Poisson statistic is no longer well-defined. This happens very rarely in this analysis, and only at energies above $\sim$100~GeV. These bins contribute negligibly to the total limit. To address these issues, we adopt the following two-prong approach. If the number of observed counts in the energy and angular bin is non-zero, we set the resulting log-likelihood to infinity. This is straightforward, since the log-likelihood expectation of producing 0 counts when $\geq$1 count is observed is also infinity. If the number of observed counts in the bin is instead 0, we set the number of predicted accounts to be equal to its absolute value. This is sensible, as it produces a best-fit point at the prediction of 0 counts. We note that these issues contribute very negligibly to the calculated limits, as energy bins above 100~GeV contribute negligibly compared to the lower-energy photon data. 

Finally, while the techniques described in this letter are widely applicable to investigations of low-significance excesses throughout the Fermi-LAT data, their interpretation may be tricky in scenarios where they are applied to studies of Fermi-LAT blazars. This is due to the fact that blazars are responsible for a significant fraction of the low-significance (TS$\sim$5) excesses observed in the Fermi-LAT data.  Thus the calculation of P$_{bg}$ will include substantial contributions from the source population it is attempting to constrain, and will associate these sources with background fluctuations. This is unlikely to affect other source classes, such as star-forming galaxies (as studied in L16), or dSphs, though we note that dark matter annihilation from subhalos can potentially produce a few percent of the underlying low-statistics fluctuations~\citep{Berlin:2013dva}. We note that in the case of blazars, variations in the standard analysis may be utilized to chose blank sky locations that significantly differ from the class of blazars being analyzed in the stacking analysis -- but care must be employed when utilizing this method.

\end{document}